\documentclass[12pt,preprint]{emulateapj}
\usepackage{amsmath}
\DeclareMathAlphabet\mathbfcal{OMS}{cmsy}{b}{n}

\usepackage{xspace,graphics,eso-pic,color,graphicx,type1cm,ulem,amsmath,color}

\shorttitle{WISE 1405}
\shortauthors{Cushing et al.}

\newcommand\teff{\mbox{$T_\mathrm{eff}$}}
\newcommand\fsed{\mbox{$f_\mathrm{sed}$}}
\newcommand\logg{\mbox{$\log g$}}

\newcommand{\wisefull}{\mbox{WISE J140518.39$+$553421.3}}
\newcommand{\wise}{\mbox{WISE J1405$+$5534}}

\newcommand{\cho}{[3.6]}
\newcommand{\cht}{[4.5]}

\begin{document}


\title{The First Detection of Photometric Variability in a Y
    Dwarf: \wisefull}


\author{Michael C. Cushing\altaffilmark{1}, Kevin
  K. Hardegree-Ullman\altaffilmark{1}, Jesica L. Trucks\altaffilmark{1},
  Caroline V. Morley\altaffilmark{2}, John E. Gizis\altaffilmark{3},
  Mark S. Marley\altaffilmark{4}, Jonathan J. Fortney\altaffilmark{2},
  J. Davy Kirkpatrick\altaffilmark{5}, Christopher
  R. Gelino\altaffilmark{5}, Gregory N. Mace\altaffilmark{6}, Sean
  J. Carey\altaffilmark{7}}

\altaffiltext{1}{The University of Toledo, 2801 West Bancroft Street,
  Mailstop 111, Toledo, OH 43606, USA; michael.cushing@utoledo.edu}

\altaffiltext{2}{Department of Astronomy and Astrophysics, University of
  California, Santa Cruz, CA 95064, USA} 

\altaffiltext{3}{Department of Physics and Astronomy, University of
  Delaware, Newark, DE 19716, USA}

\altaffiltext{5}{Infrared Processing and Analysis Center, California
  Institute of Technology, Pasadena, CA 91125, USA}

\altaffiltext{4}{NASA Ames Research Center, Moffett Field, CA 94035,
  USA}

\altaffiltext{6}{Department of Astronomy, The University of Texas,
  Austin, TX 78712, USA}

\altaffiltext{7}{Spitzer Science Center, California Institute of
  Technology, Pasadena, CA 91125, USA}

\begin{abstract}

  We present the first detection of photometric variability of a spectroscopically-confirmed Y dwarf.  The Infrared Array Camera on board the \textit{Spitzer Space Telescope} was used to obtain times series photometry at 3.6 and 4.5 $\mu$m over a twenty four hour period at two different epochs separated by 149 days.  Variability is evident at 4.5 $\mu$m in the first epoch and at 3.6 and 4.5 $\mu$m in the second epoch which suggests that the underlying cause or causes of this variability change on the timescales of months.  The second-epoch [3.6] and [4.5] light curves are nearly sinusoidal in form, in phase, have periods of roughly 8.5 hours, and have semi-amplitudes of 3.5\%. We find that a simple geometric spot model with a single bright spot reproduces these observations well.  We also compare our measured semi-amplitudes of the second epoch light curves to predictions of the static, one-dimensional, partly cloudy and hot spot models of Morley and collaborators and find that neither set of models can reproduce the observed [3.6] and[4.5] semi-amplitudes simultaneously.  More advanced two- or three-dimensional models that include time-dependent phenomena like vertical mixing, cloud formation, and thermal relaxation are therefore sorely needed in order to properly interpret our observations.

\end{abstract}

\section{Introduction}

Y dwarfs are the coolest class of brown dwarfs known \citep{cushing1,2012ApJ...753..156K} with estimated effective temperatures (\teff) below 500 K \citep[e.g.,][]{2013AAS...22115830D}. At such low temperatures, their photospheres are composed of H$_2$, He, H$_2$S, CH$_4$, H$_2$O, and NH$_3$ in the gas phase, and salt (KCl), sulfide (MnS, Na$_2$S, and ZnS), and possible water ice condensates in the solid phase which gravitationally settle within the atmosphere to form clouds \citep{2012ApJ...756..172M,2014ApJ...787...78M}.  Y dwarfs are ideal analogs to the cool gas giant exoplanets predicted to be discovered by high contrast imagers such as the Gemini Planet Imager \citep[GPI;][]{2011PASP..123..692M} and the Spectro-Polarimetric High-contrast Exoplanet REsearch \citep[SPHERE;][]{2008SPIE.7014E..18B} instrument for the Very Large Telescope.  However there are only twenty-one spectroscopically (and two photometrically) confirmed Y dwarfs known and since they are intrinsically very faint with $M_J \gtrsim 20$ mag, our understanding of their basic properties is still very limited.

The formation and subsequent evolution of condensate clouds plays a critical role in the evolution of all brown dwarfs. As a brown dwarf cools and passes through the MLTY sequence, various solid- or liquid-phase condensate clouds form until the atmosphere is eventually composed of layers upon layers of clouds, similar to that seen in Jupiter.  It has long been thought that inhomogenous cloud coverage might give rise to variations in the integrated intensity of a brown dwarf as it rotates \citep{2001ApJ...556..872A,2002ApJ...571L.151B}. Photometric variability was originally detected in L dwarfs in the $I$ band by \cite{bailer-jones1} and \cite{gelino1} but has now been detected in both L and T dwarfs, at red-optical, near-infrared, and mid-infrared wavelengths and both photometrically and spectroscopically \citep[e.g.,][]{enoch1,2006ApJ...653.1454M,buenzli1,khandrika1,2014ApJ...793...75R,2015ApJ...799..154M}. Interpretations of this variability range from simple holes in the clouds to variations in the thickness of uniform cloud decks. More recently, the potential for so-called ``hot spots'' or temperature variations giving rise to these variations has also been suggested \citep{showman1,morley2,robinson1}.

Although variability is conspicuous in L and T dwarfs, no Y dwarf variability has been reported in the literature.  We therefore initiated a Cycle 11 \textit{Spitzer Space Telescope} Exploration Science program (90015, PI: Cushing) to search for photometric variability in Y dwarfs using the Infrared Array Camera \citep[IRAC; ][]{fazio1} at $3.6\ \mu$m and $4.5\ \mu$m (hereafter \cho\ and \cht).  Fourteen Y dwarfs were observed continuously for roughly 24 hours; 12 hours at \cho\ followed by 12 hours at \cht. The observations were then repeated a few months later in order to search for any changes in the light curves.  In this work, we present the detection of photometric variability in the Y0.5pec?\footnote{\wise\ was classified as Y0p?  by \citet{cushing1} because the peak of the $H$-band emission was shifted redward by roughly 60 \AA\ relative to other late-type T dwarfs and early-type Y dwarfs. \citet{2015ApJ...804...92S} later reclassified \wise\ as Y0.5 but dropped the pec? by mistake.}  dwarf \wisefull\, \citep[hereafter \wise ;][]{cushing1}.  A future paper will discuss the results from the entire survey. 

\begin{figure*}
\includegraphics[angle=0, width=40pc]{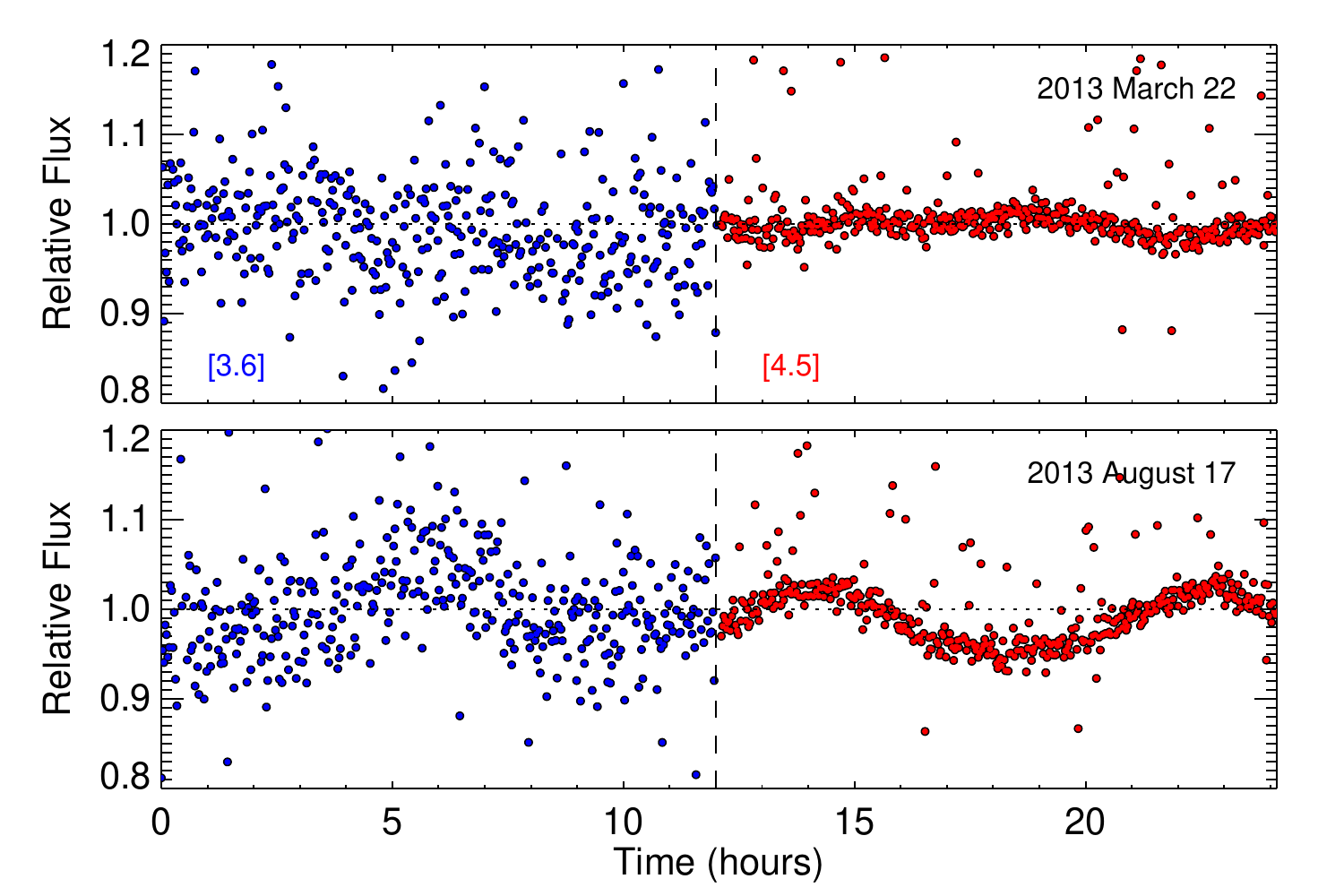}
\caption{Normalized IRAC \cho\ and \cht\ photometry on 2013 Mar 22 (top)
  and 2013 Aug 17 (bottom) plotted from the beginning of \cho\
  observations in each epoch on the same time scale. The
    ordinate range was selected to emphasize the variability in the
    light curves and thus some outlier data points with relative fluxes
    outside of this range are not shown.  The scatter in the [3.6] data
  is larger than that in the [4.5] data because \wise\ is fainter by a
  factor of seven at 3.6 $\mu$m due to the strong $\nu_3$ fundamental
  band of CH$_4$ centered at 3.3 $\mu$m.}
     \label{fig:2panel}
\end{figure*}

\section{\textit{Spitzer} Observations and Data Reduction}

\wise\ was observed on 2013 March 22 (hereafter epoch 1) and 2013 August 17 (hereafter epoch 2) with IRAC at both \cho\ and \cht.  The \textit{Spitzer} Astronomical Observation Request (AOR) numbers for these observations are 47166208, 47162624, 47173888, and 47169792.  It was observed in ``staring mode'' whereby a target is held close to the same position on the array in order to minimize the effects that variations in the quantum efficiency across an individual pixel has on the resulting photometry \citep[i.e. the pixel phase effect;][]{reach1}. \wise\ was also positioned in the upper left corner of the detector at the ``sweet spot'' pixel so that we could use the high resolution gain map of the sweet spot.  During each epoch, a series of roughly 430 images, each with an exposure time of 100 sec, were obtained over a continuous twelve hour period in \cho\ followed by 12 hours in \cht.  Images were also taken for thirty minutes prior to the start of the two \cho\ twelve hour sequences in order to mitigate initial telescope drift and allow for the spacecraft to settle.  These data were ignored in our analysis.

We began our analysis using the Basic Calibrated Data (BCD) frames generated by the \textit{Spitzer} Science Center using version S19.1.0 of the IRAC science pipeline.  Before performing aperture photometry on the BCD images, we converted the images from units of MJy sr$^{-1}$ to total electrons by dividing the images by the flux conversion factor (the \texttt{FLUXCONV} FITS keyword) in units of MJy sr$^{-1}$/DN s$^{-1}$, multiplying by the gain in units of electrons DN$^{-1}$, and then multiplying by the exposure time in seconds.  The remainder of the reduction was performed using custom Interactive Data Language (IDL) code.  The position of \wise\ was determined using the \texttt{box\_centroider} routine with an aperture radius of 3 pixels and background annulus between 3 and 7 pixels. Using the IDL \texttt{aper} routine with the \texttt{EXACT} keyword set we measured the total number of electrons detected within a 3 pixel radius of the centroid position. Background subtraction was accomplished by computing the mean number of electrons in an annulus between 3 and 7 pixels from the target centroid. We were unable to apply the high-resolution gain map of the sweet spot pixel because \wise\ was on average 0.25 pixels away from the spot and the map is only 0.5$\times$0.5-pixel in size.  Furthermore, we did not apply the pixel phase correction using the IDL routine \texttt{pixel\_phase\_correct\_gauss} provided by the \textit{Spitzer} Science Center because it did not reduce the scatter in the data.  An alternative method to correct for the pixel phase effect used by \citet{knutson1} and \citet{heinze1} was also tested but it too did not reduce the scatter in the data.  From these tests, we can conclude that the pixel phase corrections are not significant above the random noise in our data, probably due to the faintness of our target. Finally, data points that are clear outliers (i.e. the total number of electrons exceeded the median intensity level by more than fifty times the median absolute deviation\footnote{For a data set $\boldsymbol{X}$=\{$x_1$, $x_2$, ..., $x_N$\}, the median absolute deviation (MAD) is given by MAD=median($|$ $x_i$ - median($\boldsymbol{X}$)$|$).}) were removed; this resulted in the removal of eight/three points in the epoch one [3.6]/[4.5] light curves and ten/three points in the epoch two [3.6]/[4.5] light curves.

The final time series in both epochs after dividing by the median intensity level is shown in Figure \ref{fig:2panel}.  The scatter in the [3.6] data is larger than that in the [4.5] data because \wise\ is fainter by a factor of seven at 3.6 $\mu$m due to the strong $\nu_3$ fundamental band of CH$_4$ centered at 3.3 $\mu$m.

\section{Analysis}

\subsection{Characterizing the Variability}
\label{sec:characterize}

Visual inspection of Figure \ref{fig:2panel} shows that \wise\, is variable with semi-amplitudes of a few percent at both [3.6] and [4.5]. This is the first detection of variability in a spectroscopically confirmed Y dwarf and suggests that the mid-infrared variability observed in late-type T dwarfs by \citet{2015ApJ...799..154M} continues across the T/Y boundary.  There are, however, obvious differences in the light curves between the two epochs including the lack of clear variability at [3.6] in epoch 1, the near-sinusodial shape of the [3.6] and [4.5] epoch 2 light curves, and the more complex shape (i.e. not purely sinusoidal) of the [4.5] epoch 1 light curve.  These differences indicate that the underlying cause or causes of the observed variability in \wise\ evolves on timescales of months.  Finally, although clear semi-periodic or periodic variability is detected at the few percent level in the relative light curves, the average [3.6]$-$[4.5] color shows no change between the two epochs because the average flux levels at [3.6] and [4.5] are equal within the (Poisson) uncertainties.

In order to measure the amplitude, phase, and period of the roughly sinusoidal epoch 2 light curves, we assume the data are generated from the following probabilistic model

\begin{equation}
D_i = C + A \sin \left (\frac{2\pi}{P}t_i + \phi \right ) + \epsilon,
\end{equation}

\noindent
where $D_i$ is a random variable for the number of electrons detected at the $i$th time $t_i$, $C$ is an additive constant, $A$ is the semi-amplitude, $P$ is the period, $\phi$ is the phase, and $\epsilon$ is a random variable that accounts for measurement error that has a mean of zero and a variance of $\sigma^2$.  We can determine the joint probability distribution function of these five parameters given our $N$ observations (denoted as $\boldsymbol d=\{d_1, d_2, d_3, \cdots, d_N$\}) using Bayes' Theorem

\begin{equation}
 p(C, A, P, \phi, \sigma| \boldsymbol d) \,\, \propto \,\,\mathcal{L} (\boldsymbol d|C, A, P, \phi, \sigma)\,\, p(C,A,P,\phi,\sigma),
\end{equation}

\noindent
where $p(C, A, P, \phi, \sigma| \boldsymbol d)$ is the posterior distribution, $\mathcal{L}(\boldsymbol d|C, A, P, \phi, \sigma)$ is the likelihood, and $p(C,A,P,\phi,\sigma)$ is the prior distribution.  We assume that the random variable $\epsilon$ follows a normal distribution and thus assuming the data points are independent, the likelihood is given by,

\small
\begin{align}
\mathcal{L}(\boldsymbol d|C, A, P, \phi, \sigma) =  & \left ( \frac{1}{\sqrt{2\pi \sigma^2}} \right )^N \\
& \text{exp} \left [ - \sum_{i=0}^N \left ( \frac{[d_i - C - A\sin (2\pi t_i/P + \phi)]^2 }{2\sigma^2} \right ) \right ] .
\end{align}
\normalsize

Visual inspection of Figure \ref{fig:2panel} clearly shows outliers in the data.  We model the entire time series following the procedure described in \citet{2010arXiv1008.4686H} whereby we assume that the good data points are generated from Equation 1 and the outliers (i.e. ``bad'' data points) are generated from a normal distribution with a mean of $Y_\text{bad}$ and a variance of $\sigma_\text{bad}^2$.  The likelihood function then becomes,

\small
\begin{align}
\mathcal{L}(\boldsymbol d|P_\text{b}, Y_\text{bad}, \sigma_\text{bad}, C, A, P, \phi, \sigma) = & \\
 \prod_{i=0}^N \left [ \frac{1-P_\text{bad}}{\sigma \sqrt{2\pi} }\,\, \text{exp} \left ( -\frac{[d_i - C - A\sin (2\pi t_i/P + \phi)]^2}{2\sigma^2}\right ) \right. \nonumber \\
              \left. + \frac{P_\text{bad}}{\sqrt{2\pi \sigma_\text{bad}^2}}\,\, \text{exp} \left ( - \frac{[d_i - Y_\text{bad}]^2}{2\sigma_\text{bad}^2}\right ) \right ]
\end{align}
\normalsize

\noindent
where $P_\text{b}$ is the probability that a data point is bad (see \citeauthor{2010arXiv1008.4686H} for a derivation of this equation).

\begin{figure*}
\begin{center}
\includegraphics[angle=0, width=40pc]{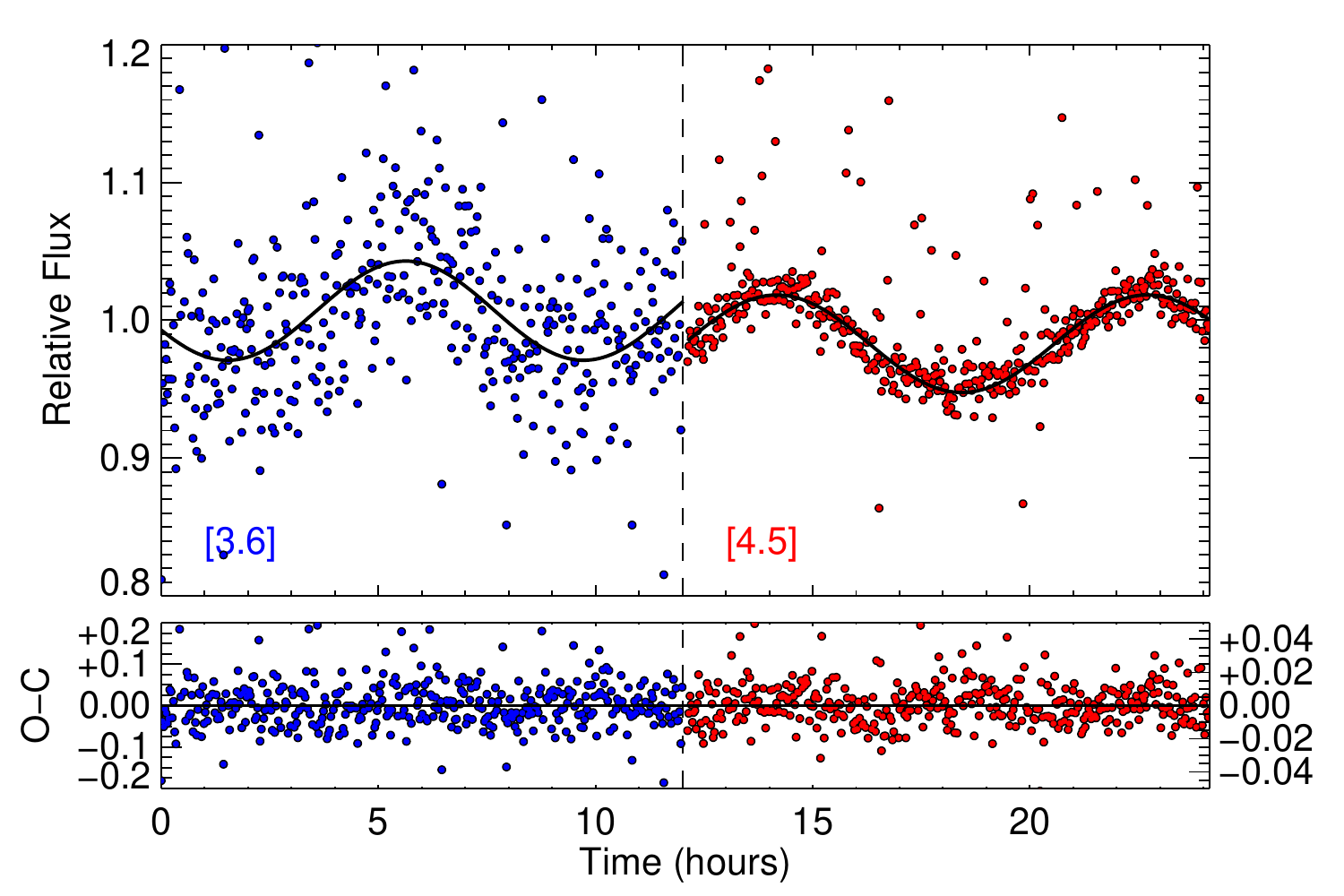}
\caption{\textit{Top:} Best fitting sine curve overplotted on the epoch
  2 [3.6] and [4.5] data.  \textit{Bottom:} The residuals ($O-C$) show
  no large-scale structure, indicating a sine curve is a reasonable
  model of the data.}
     \label{fig:sinmcmcfit}
  \end{center}
\end{figure*}

We assume that the joint prior distribution can be factored as the product of individual probability distribution functions and use uniform priors for all parameters (see Table \ref{tab:sinresults}).  We sampled the joint posterior distribution using a Markov Chain Monte Carlo method employed by the \texttt{emcee} package \citep{2013PASP..125..306F}.  We use 1000 walkers in the eight-dimensional parameter space to model the light curves and kept 700,000 samples after discarding an initial burn-in sample.  With the joint posterior distribution in hand, we computed posterior distributions for each of the eight model parameters by marginalizing over the other seven parameters.  The values corresponding to the 16th, 50th, and 84th percentiles of the marginalized distributions are given in Table \ref{tab:sinresults} and the resulting best fit models and residuals ($O-C$) are shown in Figure \ref{fig:sinmcmcfit}.

The semi-amplitudes, periods, and phases of the [3.6] and [4.5] epoch 2 light curves are equal within the uncertainties.  Since the [4.5] period has a smaller uncertainty, we identify the rotation period of \wise\ based on the [4.5] data as 8.54$\pm$0.08 hours.  The uncertainties at [3.6] and [4.5] of 0.045 and 0.011 (45 mmag and 11 mmag) are factors of $\sim$3 and $\sim$2 larger than the theoretical photon limit.  The residuals also show no large-scale trends, which suggests that a simple sine curve is an accurate representation of the data.

\begin{deluxetable*}{llcc}
\tablecolumns{3}
\tabletypesize{\scriptsize} 
\tablewidth{0pc}
\tablecaption{\label{tab:sinresults}Epoch 2 Sine Curve Model Parameters}
\tablehead{
\colhead{Model Parameter} & 
\colhead{Prior\tablenotemark{a}} & 
\colhead{[3.6] Value\tablenotemark{b}} & 
\colhead{[4.5] Value}}

\startdata
Constant $C$                                    & $\mathcal{U}$(0.9,1.1)   & 1.007$\pm$0.003         & 0.9834$\pm$0.0006 \\
Amplitude $A$ (\%)                              & $\mathcal{U}$(0,5)       & 3.6$\pm0.4$             & 3.54$\pm$0.09  \\ 
Period $P$ (hours)                              & $\mathcal{U}$(7,10)      & $8.2\pm0.3$             & 8.54$\pm$0.08 \\
Phase $\phi$ (degrees)                          & $\mathcal{U}$(0,360)     & 203$^{+12}_{-11}$       & 213$\pm$7   \\
Standard deviation $\sigma$                     & $\mathcal{U}$(0,0.5)     & 0.045$\pm$0.002         & 0.0109$\pm$0.0005 \\
Bad data mean $Y_\text{bad}$                    & $\mathcal{U}$(0.5,2.5)   & 1.44$^{+0.09}_{-0.08}$  & 1.07$\pm$0.02  \\ 
Bad data standard deviation $\sigma_\text{bad}$ & $\mathcal{U}$(0,1)       & 0.50$^{+0.6}_{-0.05}$   & 0.11$\pm$0.01 \\ 
Bad data probability $P_\text{bad}$             & $\mathcal{U}$(0,0.25)    & 0.10$\pm$0.02           & 0.12$\pm$0.02 \\ 
\enddata
\tablenotetext{a}{$\mathcal{U}(a,b)$ denotes a uniform distribution over the
  range $a$ to $b$.}

\tablenotetext{b}{The values reported correspond to the 16th, 50th, and
  84th percentiles of the marginalized posterior distribution.}

\end{deluxetable*}

\subsection{Interpreting the Variability}

There are several mechanisms that can induce variability in the integrated light of low-mass stars and brown dwarfs including magnetic activity, non-uniform surface opacities, either in the form of chemical abundance variations or heterogeneous cloud coverage, and non-uniform temperature profile resulting in ``hot'' and/or ``cold'' spots. Magnetic activity is often ignored in the study of cool brown dwarfs because the atmospheres of brown dwarfs are predominantly neutral and thus the magnetic field lines presumably have a difficult time coupling to the gas \citep[e.g.,][]{gelino1,2002ApJ...571..469M}.  The degree to which non-uniform surface opacities and non-uniform temperature profiles contribute to the observed levels of variability in L and T dwarfs and the timescales on which they operate are still an open question, although some progress has been made.

\begin{figure} 
\centerline{\hbox{\includegraphics[width=3.5in,angle=0]{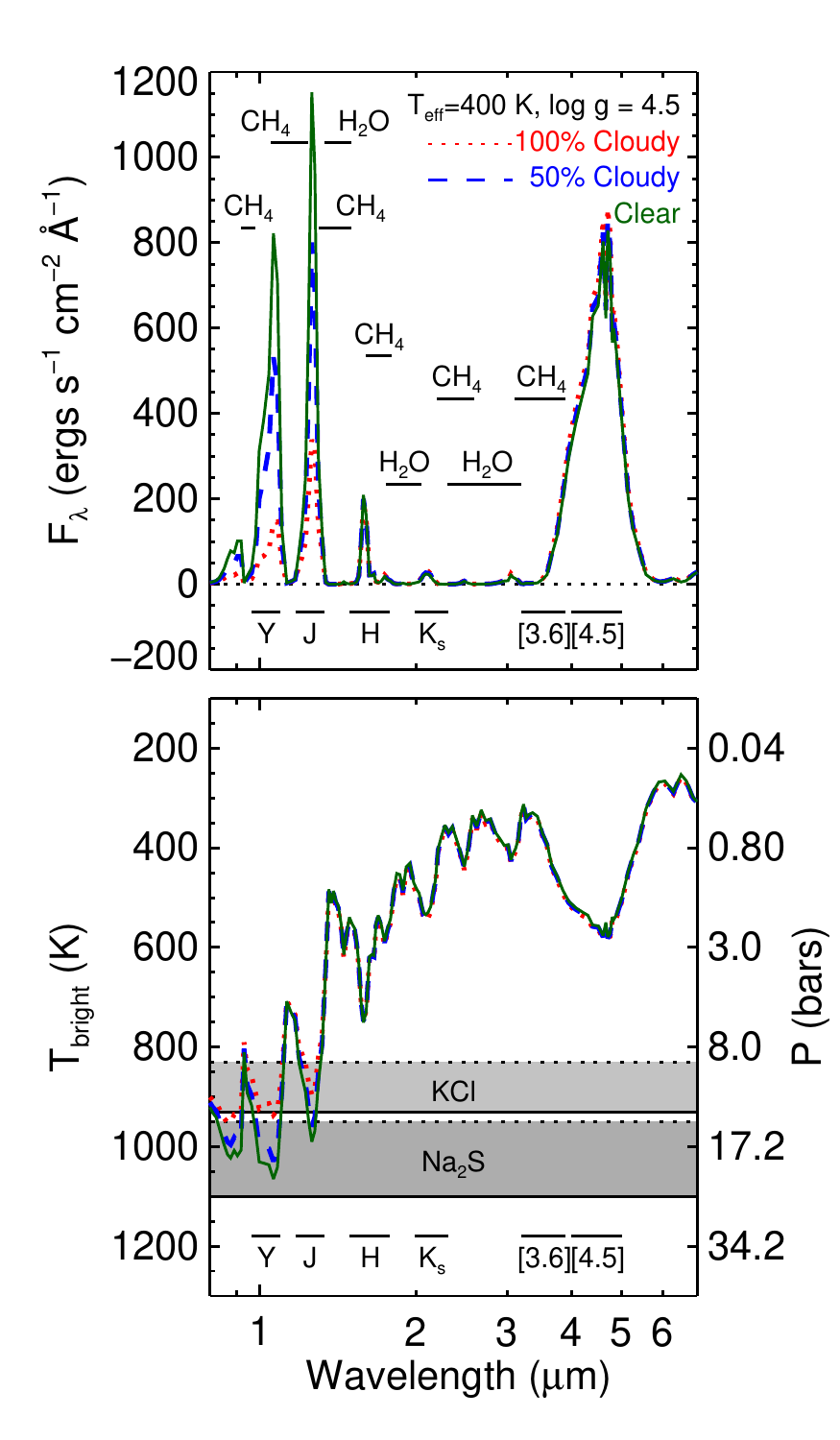}}}
\caption{\label{fig:TP}\textit{Top:} Solar metallicity partly cloudy
  model spectra from \cite{morley2} at \teff=400 K, \logg = 4.5 (cm
  s$^{-2}$), \fsed=3 for three different cloud coverages, 100\% cloudy,
  50\% cloud, and clear.  Prominent molecular absorption bands are
  identified along with the wavelength range between the half-power
  points of the $Y$ band \citep{2002PASP..114..708H}, the 2MASS JHK$_s$
  bands \citep{2003AJ....126.1090C}, and the IRAC [3.6] and [4.5]
  bandpasses \citep{fazio1}.  \textit{Bottom:} Brightness temperature
  for the same models.  The locations of the Na$_2$S and KCl cloud decks
  are indicated.  The pressure corresponding to the brightness
  temperature derived using the P/T profile of the clear model is also
  given.}
\end{figure}

\subsubsection{Vertical Extent of Spot(s)}
\label{sec:extent}

Before comparing our observations to the detailed predictions of model atmospheres, we first discuss what can be learned about the underlying cause of the variability by exploiting the near-simultaneous, multi-wavelength nature of the observations.  The emergent spectra of brown dwarfs are distinctly non-Planckian and as a result, different wavelengths probe different layers of the atmosphere.  Detecting and characterizing variability at multiple wavelengths simultaneously can therefore inform our understanding of the vertical extent of the underlying cause (or causes) of the variability.  As an example, the top panel of Figure \ref{fig:TP} shows the emergent intensity for three atmospheric models from \citet{morley2} with \teff=400 K, \logg=4.5 (cm s$^{-2}$) , \fsed=3 \footnote{The sedimentation efficiency parameter describes the efficiency of cloud particle sedimentation relative to turbulent mixing \citep{2001ApJ...556..872A}.  Larger values of \fsed\ imply larger particle sizes and geometrically thinner clouds. Cloudless models are denoted as \fsed=nc where nc means ``no clouds''.}  and cloud coverage fractions of 0\%, 50\%, and 100\% (h=1, 0.5, 0, see \S\ref{sec:partlycloudy} and Equation \ref{eq:holes}).  The lower panel of Figure \ref{fig:TP} shows the brightness temperature of the same three models.  The brightness temperature is the temperature of a blackbody that has the same intensity as the model at a given wavelength.  Since temperature typically increases with increasing depth into an atmosphere, the brightness temperature can be used as a proxy for depth in the atmosphere.  Also given are the atmospheric pressures derived using the model pressure--temperature profile that corresponds to the brightness temperatures.

The red optical (0.70--1.0 $\mu$m) through mid-infrared wavelengths probe a large range of atmospheric temperatures and pressures.  In general, the [3.6] and [4.5] IRAC bands probe higher atmospheric layers (i.e. lower temperatures, lower pressures) than near-infrared wavelengths (1--2.5 $\mu$m).  However, due to the strong methane absorption band centered at 3.3 $\mu$m, the [3.6] band probes atmospheric layers with $P$ $\approx$ 0.8 bar and $T$ $\approx$ 400 K while the [4.5] band probes layers with $P$ $\approx$ 3 bar and $T$ $\approx$ 600 K.  Although the brightness temperature is a useful proxy for the pressure level at which radiation emerges at a given wavelength, multiple layers of the atmosphere actually contribute to the emergent flux at each wavelength.  In addition, the \textit{Spitzer} bandpasses are not delta functions and thus the radiation detected through them actually emerges from an even broader range of pressure levels.

\begin{figure*}[bh]
\centerline{\hbox{\includegraphics[width=6in,angle=0]{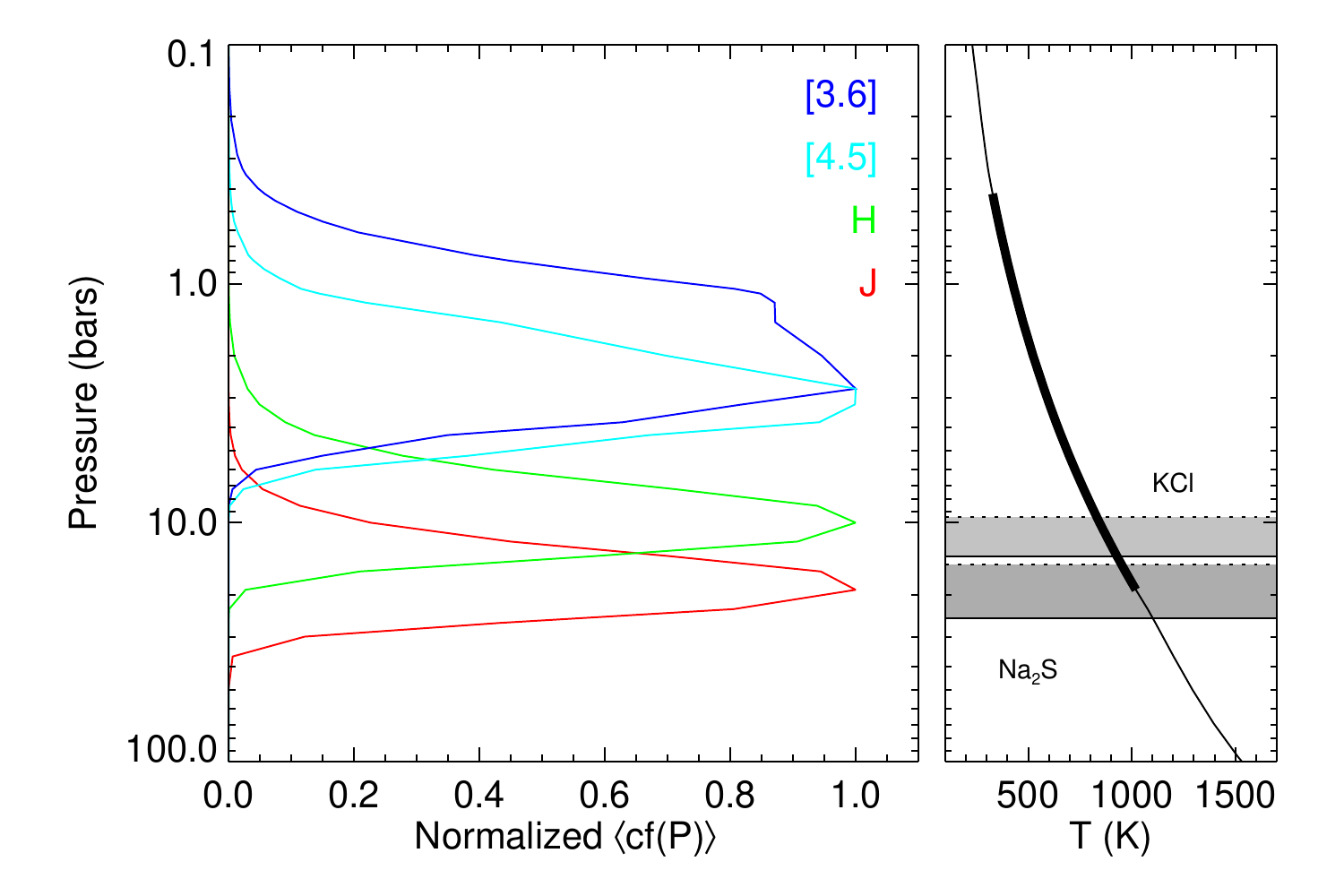}}}
\caption{\label{fig:CF}\textit{Left:} Band-averaged
    contribution functions normalized to unity at their peak for the
    $J$, $H$, and [3.6] and [4.5] bandpasses for a cloudless model
    atmosphere with \teff=400 K and \logg=4.5 (cm s$^{-2}$).
    \textit{Right:} The temperature/pressure profile for the same model
    with the location of the KCl and Na$_2$S clouds indicated.  The
    thick line denotes the region of the atmosphere that is convective.} 
\end{figure*}

A more accurate representation of what pressure levels contribute \textit{thermal emission} to the emergent flux at a given wavelength is given by the contribution function \citep{1987IGS....36.....C},

\begin{equation}
  \text{cf}(\lambda,P) = B_\nu(\lambda, T ) \, e^{-\tau_\lambda} \frac{d\tau_\lambda}{d\log P},
\end{equation}

\noindent
where $B_\nu(\lambda, T)$ is the Planck function, $P$ is the atmospheric pressure, and $\tau_\lambda$ is the \textit{vertical} monochromatic optical depth; the integral of the contribution function over pressure in a semi-infinite atmosphere gives the specific intensity at the top of the atmosphere,

\begin{equation}
I_\lambda (\lambda, P=0) =  \int_\infty^0 \text{cf}(\lambda,P)\,\, d\log P.
\end{equation}

\noindent
Since the [3.6] and [4.5] bandpasses have a finite width, we must compute a band-averaged contribution function which is given by the integral of the contribution function over the system response function $S(\lambda)$,

\begin{equation}
  \langle \text{cf}(P) \rangle = \frac{\int_0^\infty \text{cf}(\lambda,P)\, S(\lambda)\, d\lambda} {\int_0^\infty S(\lambda) d\lambda },
\end{equation}

\noindent
where the integral over the band-averaged contribution function is now proportional to flux detected through the bandpass.

We computed contribution functions at two thousand wavelengths and sixty pressure levels within a \textit{cloudless} model atmosphere with \teff=400 K, \logg=4.5 [cm s$^{-2}$] and then computed the band-averaged contribution functions for the $J$, $H$, [3.6], and [4.5] bands using the full array average system response functions of the [3.6] and [4.5] bands\footnote{Files \texttt{080924ch1trans\_full.txt} and \texttt{080924ch2trans\_full.txt} obtained on 2014 May 22 from http://irsa.ipac.caltech.edu/data/SPITZER/docs/irac/calibrationfiles/spectralresponse/.} and the Two Micron All Sky Survey (2MASS) $J$ and $H$ band transmission curves from \citet{2003AJ....126.1090C}.  Figure \ref{fig:CF} shows the resulting normalized band-averaged contribution functions for the $J$, $H$, [3.6] and [4.5] bandpasses along with the temperature-pressure profile for the cloudless \teff=400 K, \logg=4.5 [cm s$^{-2}$] model.

Although on average the [3.6] band probes slightly higher pressures than the [4.5] band, there is considerable overlap and thus much of the emergent flux in the \textit{Spitzer} bandpasses come from the same layers of the atmosphere.  Nevertheless, a few general conclusions can be still drawn.  If the variability is caused by temperature perturbations at depth, then the fact that the epoch 2 [3.6] and [4.5] light curves have the same phase within the uncertainties suggests that whatever phenomenon causes these perturbations extends over at least these atmospheric layers, e.g. 3--0.1 bars.  In contrast, the lack of variability detected at [3.6] in the epoch 1 data suggests that the phenomenon is either absent from the upper atmosphere $\lesssim$0.6 bar or has weakened in strength such that its effect on the emergent flux falls below our detection threshold.  The differences in shape of the two [4.5] light curves also suggest that changes in the phenomenon deeper in the atmosphere have also occurred between the two epochs.  If the observed variability is caused by temperature variations, then our observations provide the first evidence for the evolution of weather patterns with depth on the timescales of months for Y dwarfs.

\subsubsection{Geometric Spot Model}

Before delving into the possible underlying cause or causes of the observed variability, we first model the variability using a simple geometric spot model because it requires the least number of assumptions.  The simplicity and similarity of the \wise\ epoch 2 light curves suggests that a single feature may explain the variability.  We therefore modeled both the [3.6] and [4.5] epoch 2 light curves with a single, bright circular spot using the equations of \citet{1987ApJ...320..756D}. The brown dwarf is characterized by a rotation period, the inclination to the line-of-sight, and the brightness of the spot-free photosphere at [3.6] and [4.5].  The spot is characterized by a size (radius, in radians), position (longitude and latitude), and the spot-to-photosphere brightness ratio at [3.6] and [4.5].  This ratio is defined as the flux per unit area of the spot divided by the flux per unit area of the unspotted photosphere; it is greater than one for bright spots. Lacking constraints on the limb darkening, we adopt linear limb darking coefficients of 0.5 for both the photosphere and spot ($\mu_\ast = \mu_s= 0.5$ in Dorren's notation.)

\begin{figure*}
\begin{center}
\includegraphics[angle=0, width=40pc]{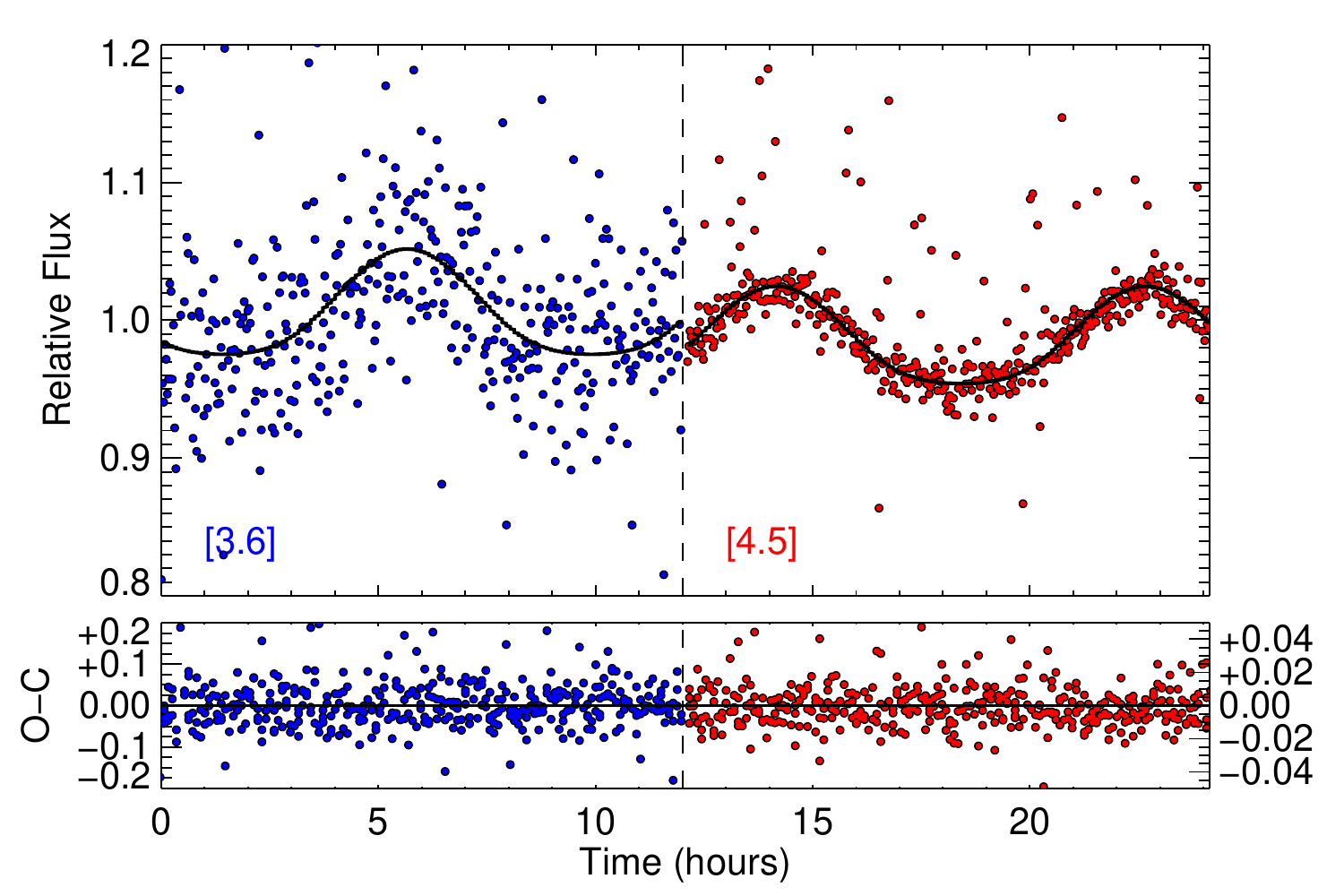}
\caption{\textit{Top:} Best fitting spot model overplotted on the epoch
  2 [3.6] and [4.5] data.  \textit{Bottom:} The residuals ($O-C$) show
  no large-scale features, which suggests the single spot model is a
  reasonable model of the data.}
     \label{fig:spotmcmcfit}
  \end{center}
\end{figure*}

To explore the possible values of the parameters, we use the \texttt{emcee} code \citep{2013PASP..125..306F} to perform a Markov Chain Monte Carlo calculation of the joint posterior distribution using the same model for bad data that is described in \S \ref{sec:characterize}.  We use 1000 walkers in the seventeen-dimensional parameter space (eleven parameters for the spot model and six parameters to account for the bad data) to model the light curves and kept 2.9 million samples after discarding an initial burn-in sample.  The values corresponding to the 16th, 50th, and 84th percentiles of the marginalized distributions of each parameter are given in Table 2.  We find that the rotation period is $8.50\pm 0.05$ hours (consistent with the period derived \S \ref{sec:characterize}), the size of the spot is $14\pm3$ degrees and the spot-to-photosphere brightness ratio at [3.6] is $4.0^{+1.2}_{-0.9}$ and at [4.5] is $3.8^{+1.0}_{-0.8}$.  Although a broad range of spot brightnesses are consistent with the data, the brightness in the two bands must be similar since the [3.6]/[4.5] ratio is $1.05\pm0.08$. A ratio of near unity is also consistent with the fact that [3.6] and [4.5] band-averaged contribution functions overlap (see \S \ref{sec:extent}) which suggests that a single spot could account for the variability in both bands. The light curves corresponding to the best fit values and residuals $(O-C)$ are shown in Figure \ref{fig:spotmcmcfit}.

\begin{deluxetable*}{llc}
\tablecolumns{3}
\tabletypesize{\scriptsize} 
\tablewidth{0pc}
\tablecaption{\label{tab:spotresults}Epoch 2 Spot Model Parameters}
\tablehead{
\colhead{Model Parameter} & 
\colhead{Prior\tablenotemark{a}} & 
\colhead{Value\tablenotemark{b}}}

\startdata
Spot radius $\alpha$ (degrees) & $\mathcal{U}(0,90)$ & $14\pm 3$ \\
Spot latitude $\chi$ (degrees) & $\mathcal{U}(0,90)$ & $13\pm 4$ \\
Stellar inclination $i$ (degrees) & $\mathcal{U}(0,90)$ & $16^{+5}_{-4}$ \\
Spot longitude $\psi$ (degrees) & $\mathcal{U}(0,360)$ & 99$\pm$7 \\
Period $P$ (hours) & $\mathcal{U}(0,19)$ & 8.50$\pm 0.05$  \\
$[3.6]$ Photosphere flux $F_{*,[3.6]}$ & $\mathcal{U}(0.5,15)$ & $0.972^{+0.005}_{-0.007}$ \\
$[3.6]$ Spot-to-photosphere flux ratio $(F_s/F_*)_{[3.6]}$ & $\mathcal{U}(1,99)$  & $4.0^{+1.2}_{-0.9}$ \\
$[4.5]$ Photosphere flux $F_{*,[4.5]}$ & $\mathcal{U}(0.5,15)$ & $0.951^{+0.003}_{-0.005}$ \\
$[4.5]$ Spot-to-photosphere flux ratio $(F_s/F_*)_{[4.5]}$ & $\mathcal{U}(1,99)$ & $3.8^{+1.0}_{-0.8}$ \\
$[3.6]$ Standard deviation $\sigma_{[3.6]}$ & $\mathcal{U}(0,0.5)$ & $0.044\pm 0.002$            \\
$[4.5]$ Standard deviation $\sigma_{[4.5]}$  & $\mathcal{U}(0,0.5)$ & $0.0102\pm 0.0005$       \\
$[3.6]$ Bad data mean $Y_{[3.6],\text{bad}}$  & $\mathcal{U}(0.5,2.5)$ & $1.4^{+0.07}_{-0.06}$         \\
$[3.6]$ Bad data standard deviation $\sigma_{[3.6],\text{bad}}$ & $\mathcal{U}(0,1)$ & $0.48^{+0.05}_{-0.04}$ \\
$[3.6]$ Bad data probability $P_{[3.6],\text{bad}}$ & $\mathcal{U}(0,0.25)$ & $0.10\pm 0.02$      \\
$[4.5]$ Bad data mean $Y_{[4.5],\text{bad}}$ & $\mathcal{U}(0.5,2.5)$ & $1.07\pm 0.01$          \\
$[4.5]$ Bad data standard deviation $\sigma_{[4.5],\text{bad}}$ & $\mathcal{U}(0,1)$ & 0.10$\pm 0.01$ \\
$[4.5]$ Bad data probability $P_{[4.5],\text{bad}}$ & $\mathcal{U}(0,0.25)$ & $0.13\pm 0.02$      

\enddata
\tablenotetext{a}{$\mathcal{U}(a,b)$ denotes a uniform distribution over the
  range $a$ to $b$.}

\tablenotetext{b}{The values reported correspond to the 16th, 50th, and
  84th percentiles of the marginalized posterior distribution.}

\end{deluxetable*}

\pagebreak

\subsubsection{Partly Cloudy Model Atmospheres}
\label{sec:partlycloudy}

We now compare our observations to the predictions of the one-dimensional model atmospheres of \citet{morley2}.  These models are computed assuming solar abundance and include salt (KCl), sulfide (MnS, Na$_2$S, and ZnS), chromium, and water ice clouds \citep{2012ApJ...756..172M,2014ApJ...787...78M}.  The model atmospheres are static and thus show no variation in intensity with time.  We can, however, compare the measured semi-amplitudes of the \wise\ epoch 2 light curves to those computed from models with different cloud coverage fractions.  Partly cloudy atmospheres are modeled following the prescription of \citet{marley1} whereby the flux through a cloudy column of gas and a clear column of gas are computed separately and then combined to calculate the total flux through a partly cloudy column as,

\begin{equation}
\label{eq:holes}
F_\text{total} = hF_\text{clear} + (1-h)F_\text{cloudy},
\end{equation}

\noindent
 where $h$ is a parameter that ranges from zero to one and gives the fraction of the atmosphere covered by holes.  In order to simulate all possible variations in $h$ from zero to one, a suite of sixty six models with \teff=400, 500 K, \logg = 4.0, 4.5, 5.0 (cm s$^{-2}$), \fsed=3, and $h$ values ranging from zero to one in steps of 0.1 was generated. These models cover the range of effective temperatures and surface gravities (\teff=370--470 K, \logg = 4.12--4.78 [cm s$^{-2}$]) computed by \citet{2013Sci...341.1492D} for \wise.  The effective temperature and surface gravity values from \citeauthor{2013Sci...341.1492D} were computed by first applying a model-derived bolometric correction to observed absolute magnitudes of \wise\ and then comparing the resulting bolometric luminosities to evolutionary models at ages of 1 and 5 Gyr. It should be noted that any conclusions we draw from these comparisons must be tempered against the fact that these models do not fit the near- to mid-infrared spectral energy distributions of the cold late-type T and Y dwarfs well \citep{2015ApJ...804...92S}.

Perhaps the simplest way to simulate brown dwarf variability using these models is to assume that one hemisphere has an $h$ value of $h_1$ and the other hemisphere has an $h$ value of $h_2$.  The rotation of the brown dwarf would then modulate the total intensity and produce a measurable difference in the integrated intensity.  For a given effective temperature and surface gravity, we computed synthetic semi-amplitudes for all possible ($h_1, h_2$) pairs by computing the average flux densities through the \textit{Spitzer} [3.6] and [4.5] following the prescription of \citet{2006ApJ...648..614C} and then computing the semi-amplitude as,

\begin{equation}
\label{eq:A}
A_\lambda=\frac{\text{max}[f_{\nu,h_1} (\lambda), f_{\nu,h_2}
(\lambda)]}{\text{average}[f_{\nu,h_1} (\lambda), f_{\nu,
  h_2}(\lambda)]}-1.
\end{equation}

\begin{figure}
\centerline{\hbox{\includegraphics[width=5in,angle=0]{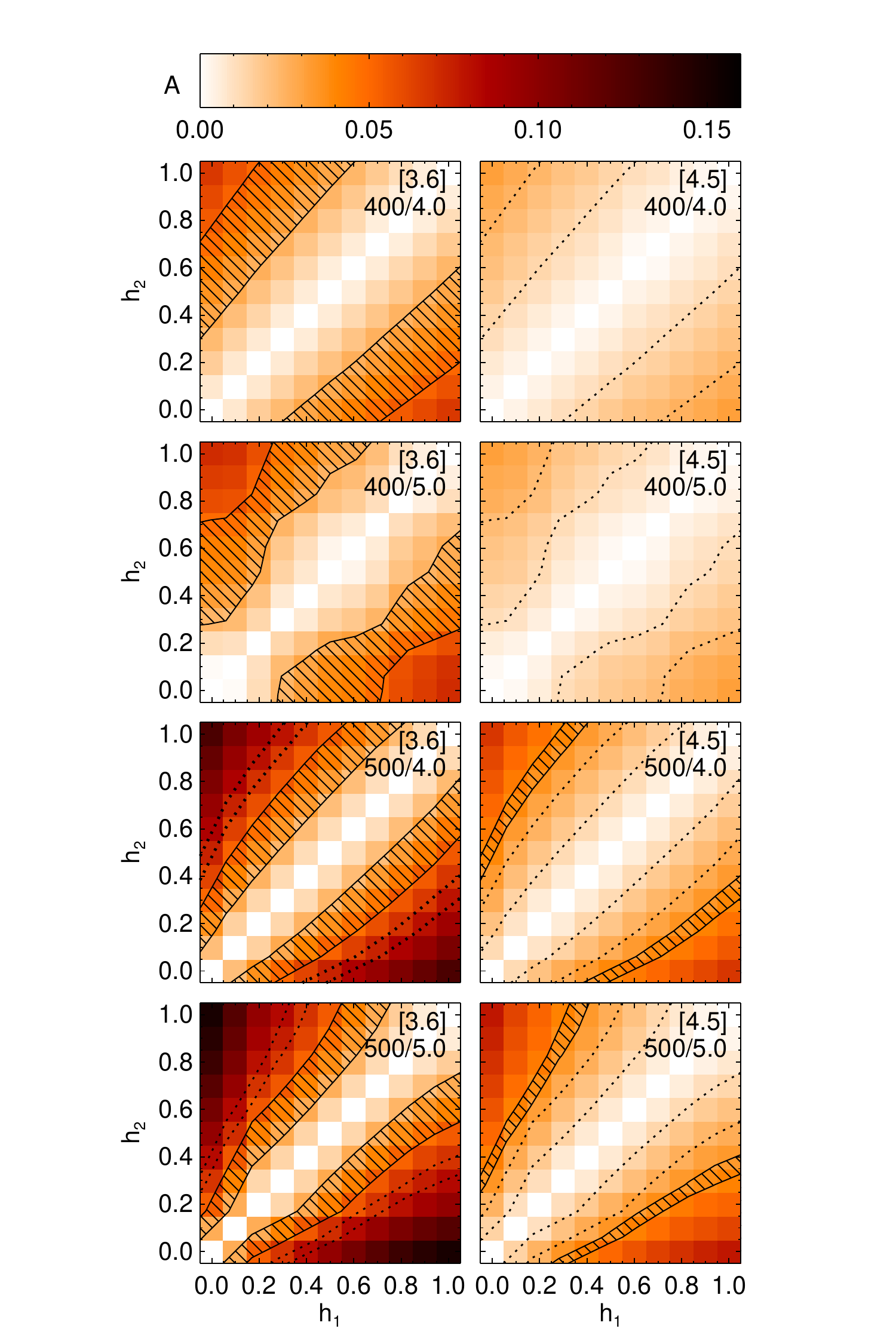}}}
\caption{\label{fig:pccomp} Semi-amplitudes at [3.6] (\textit{left
    column}) and [4.5] (\textit{right column}) arising from changes in
  the cloudiness fraction $(h_1 \rightarrow h_2)$ at different effective
  temperatures and surface gravities (\teff/\logg ).  Since
  $h_1 \rightarrow h_2$ is equivalent to $h_2 \rightarrow h_1$, each
  panel is symmetric across the diagonal.  In each panel, the
  $\pm 3\sigma$ range of the measured semi-amplitude of the second-epoch
  \wise\ light curve are shown as hashed regions while $\pm 3\sigma$
  ranges for the other wavelength are encompassed by the dotted lines.
  The model [4.5] semi-amplitudes for the 400 K models fall below the
  observational level and thus no regions are shown.  Since the contours
  never overlap, there are no pairs of $h$ values that simultaneously
  reproduce the [3.6] and [4.5] observations.}
\end{figure}

Figure \ref{fig:pccomp} shows the results for effective temperatures of 400 and 500 K and log surface gravities of 4.0 and 5.0 (cm s$^{-2}$). Several important conclusions can be drawn from this figure.  First, the models predict that the semi-amplitudes at both wavelengths are larger at \teff=500 K.  Second, variations in surface gravity at a fixed effective temperature have a small ($\lesssim$ 1\%) effect on the semi-amplitudes.  And finally, for any given ($h_1$, $h_2$) pair, the models always predict a larger semi-amplitude at [3.6].  Also shown in each panel as solid lines are the 3$\sigma$ contours corresponding to the observed epoch 2 semi-amplitudes; the 3$\sigma$ contours for the other wavelength are shown as dotted lines. Agreement between the models and observations occur if the two contours overlap since a single ($h_1, h_2$) can account for the observed semi-amplitudes at both wavelengths.  While there is no overlap in any of the panels, the two contours come close at \teff=500 K which, given the complexity of the model atmospheres, suggests that patchy clouds is a plausible mechanism to explain the variability observed of \wise\ at [3.6] and [4.5].  It should be noted, however, that a hypothetical two-hemisphere brown dwarf constructed using models with different $h$ values is unphysical since the pressure--temperature profiles of the two models do not converge to the same adiabat at depth due to the differences in cloud opacity.  In effect, we are comparing the atmospheres of two different brown dwarfs rather than the two hemispheres of a single brown dwarf.

\begin{figure}
\centerline{\hbox{\includegraphics[width=5in,angle=0]{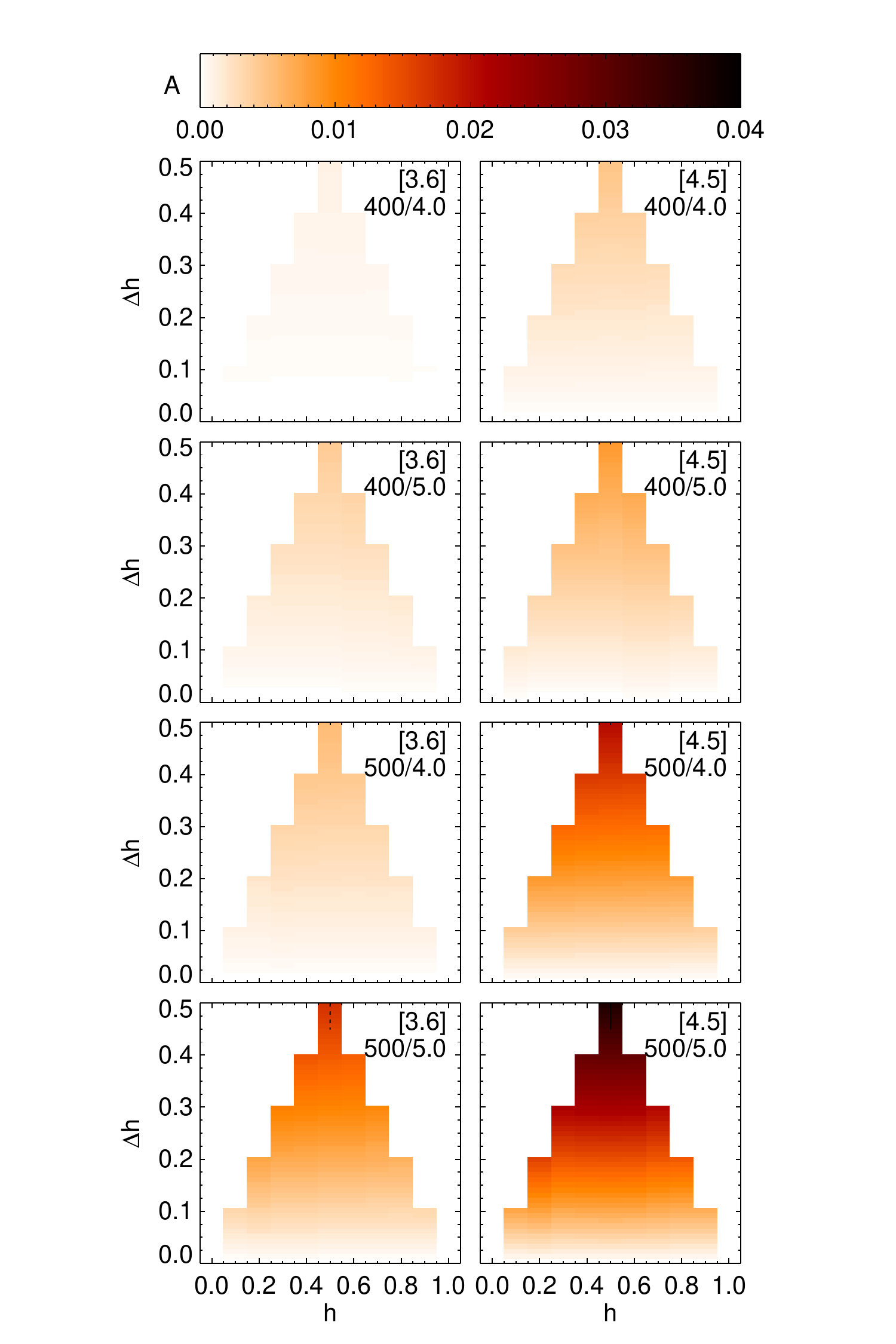}}}
\caption{\label{fig:intracomp}Semi-amplitudes at [3.6] (\textit{left
    column}) and [4.5] (\textit{right column}) arising from changes in
  the cloudiness fraction $\Delta h$ for a fixed $h$ value at different
  effective temperatures and surface gravities (\teff/\logg ).  With the
  exception of the \teff=500 K, \logg = 5.0 (cm s$^{-2}$), $h=0.5$,
  $\Delta h$=0.5 semi-amplitude, the models do not predict
  semi-amplitudes high enough to match the observations and thus regions
  similar to those shown in Figure \ref{fig:pccomp} are not indicated.}
\end{figure}

An alternative method that is self-consistent is described by \cite{morley2} whereby a model with a \textit{single} $h$ value (e.g., $h$=0.5) is used to generate two model spectra with different $h$ values (e.g., 0.4 and 0.6).  The possible combinations of $h$ values are limited, however, because the two values must always average to the $h$ value of the original model so that on the whole, the atmosphere is self consistent. To compute model spectra with $h$ values given by $h_0-\Delta h$ and $h_0+\Delta h$, we use the clear and cloudy columns for a model with an $h=h_0$ and Equation \ref{eq:holes}.  Synthetic semi-amplitudes were computed as described above and the results are shown in Figure \ref{fig:intracomp}.  Just like the previous method, the semi-amplitudes are larger at \teff=500 K but are much lower ($<$ 5\%) than those resulting from simply varying the $h$ values as described above (see also Figure \ref{fig:pccomp}).  Similarly, the models always predicted a larger semi-amplitude at [3.6] instead of [4.5].  Indeed the only model with a semi-amplitude that approaches the observations has \teff=500 K, \logg = 5.0 (cm s$^{-2}$), $h=0.5$, and $\Delta h=0.5$ and predicts $A_{[3.6]}=0.017$ and $A_{[4.5]}=0.036$.

\begin{figure}
\centerline{\hbox{\includegraphics[width=3.5in,angle=0]{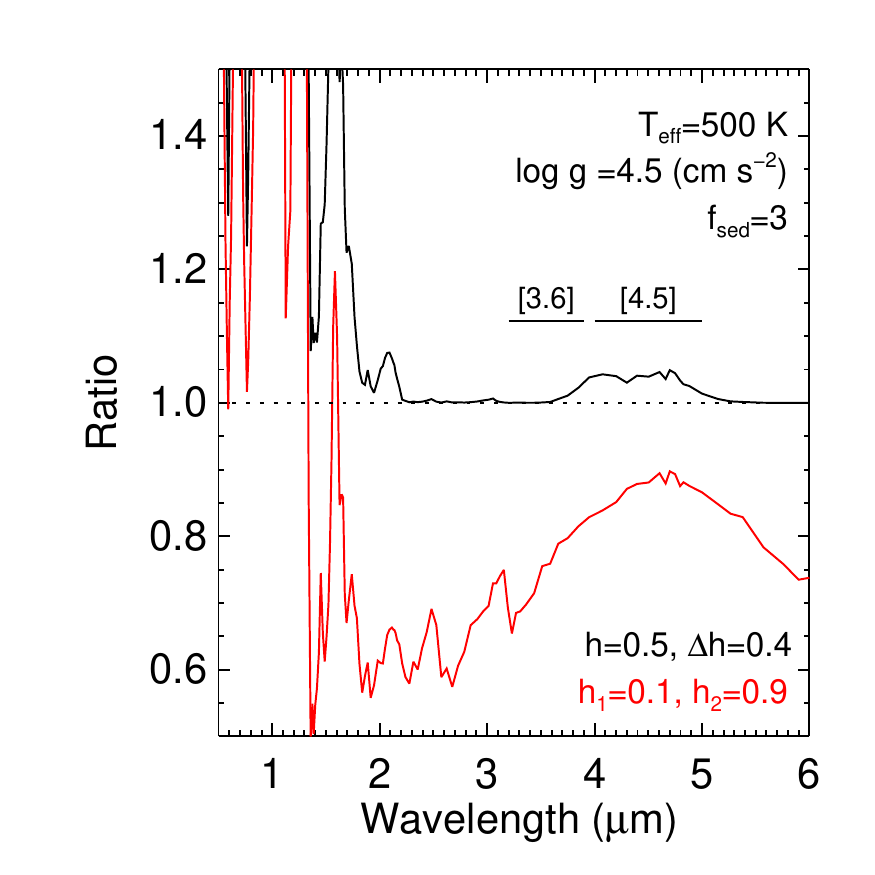}}}
\caption{\label{fig:methodcomp}Ratio of model spectra with \teff=500 K,
  \logg=4.5 (cm s$^{-2}$), \fsed=3 computed in two ways.  \textit{Red:}
  The two spectra were computed independently with hole fraction values
  of $h_1=0.1$ and $h_2=0.9$.  \textit{Black:} The two spectra were
  computed using a single hole fraction value of $h=0.5$ with
  $\Delta h=0.4$.  The 50\% power points of the IRAC [3.6] and [4.5]
  bandpasses are also indicated.}
\end{figure}

The stark contrast in the semi-amplitudes predicted by the two methods is more easily understood by plotting the ratios of the model spectra as a function of wavelength instead of focusing on the semi-amplitudes at \cho\ and [4.5].  Figure \ref{fig:methodcomp} shows the ratio of model spectra computed with $(h_1,h_2)=(0.1,0.9)$ and with $h=0.5$, $\Delta h=0.4$.  The self-consistent models with $h$ = 0.5, $\Delta h$ = 0.4 result in large ratios at near-infrared wavelengths but small ratios at mid-infrared wavelengths. For these models, the pressure-temperature profile used to calculate the spectra is the same (converged with h=0.5); the cloud opacity is the only difference between the hemispheres. This means that the differences in flux are due only to the opacity of the clouds limiting the depth from which flux emerges. In contrast, the ratio of two models computed with two different $h$ values ($h_1$,$h_2$) = (0.1,0.9) show large ratio values across all wavelengths, including decreased flux at longer wavelengths. In these models, the total flux emerging from each hemisphere is held constant (both have an effective temperature of 500 K), but the pressure-temperature profiles are different by tens of degrees. The total opacity is lower in the less cloudy model, causing the temperature profile to be cooler than the cloudy model to emit the same total flux. This means that the flux in the near-infrared is higher for the clearer model (due to the lower cloud opacity) and the flux in the mid-infrared is smaller (due to the lower temperature at a given pressure).

\subsubsection{Hot Spots}

\begin{figure} 
\centerline{\hbox{\includegraphics[width=3.5in,angle=0]{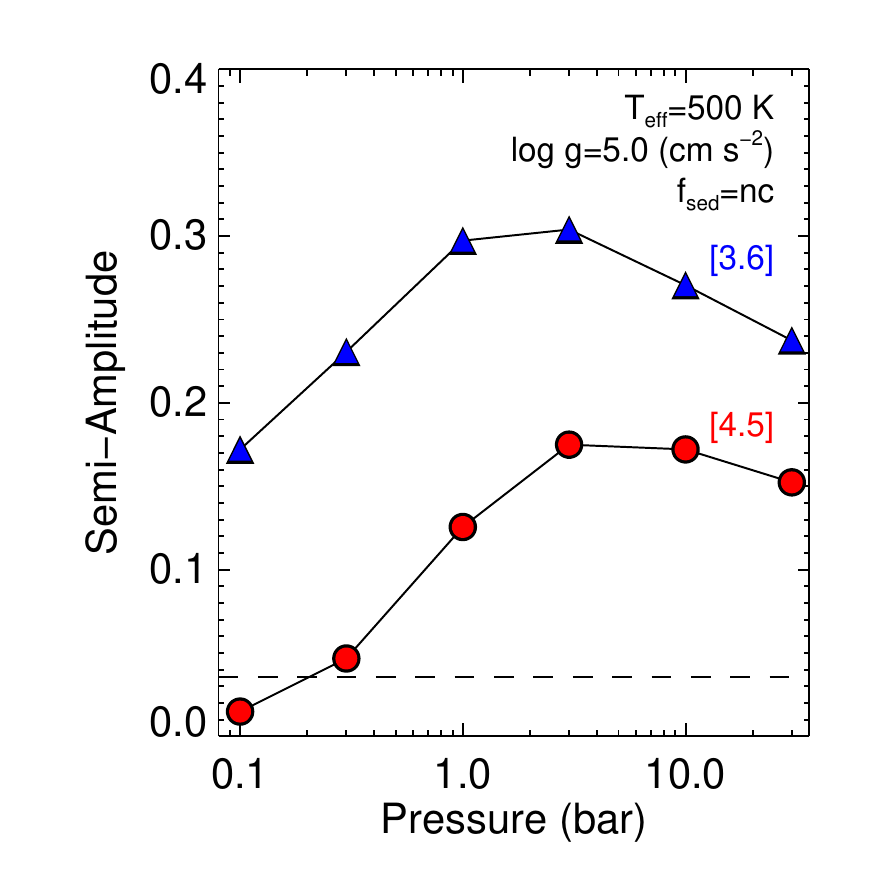}}}
\caption{\label{fig:hotspot} Semi-amplitudes at [3.6] (\textit{blue
    triangles}) and [4.5] (\text{red circles}) as a function of the
  depth of the atmosphere at which thermal energy is injected for a
  cloudless (\fsed=nc) model with \teff=500 K, \logg=5.0 (cm
  $s^{-2}$). The average of the near-equal [3.6] and [4.5]
  semi-amplitudes of the second-epoch light curve of \wise\ is shown as
  a dashed line.}
\end{figure}

Variability can also be generated by atmospheric dynamics that perturb the temperature structure of an atmosphere.  For example, the rising or sinking of air parcels on a timescale faster than the air parcels can equilibrate with their surroundings will produce ``hot'' or ``cold'' spots that could modulate the integrated intensity as the object rotates.  Vertically propagating temperature perturbations, whether induced by atmospheric waves or simple radiative coupling \citep[e.g.,][]{robinson1} could also potentially produce localized temperature anomalies. \citet{morley2} simulated the perturbations arising from hot spots by injecting energy ($F_\text{new}=1.5 F_\text{baseline}$) at various pressure levels (0.1 to 30 bars in steps of $\log_{10} (3)$ dex) into static, one-dimensional, and cloudless (\fsed=$\infty$) model atmospheres. Model spectra are then generated by assuming that 5\% of the surface is covered by the hot spot. We then determined a synthetic semi-amplitude as a function of the depth of the energy deposition by computing the average flux densities through the \textit{Spitzer} [3.6] and [4.5] bandpasses for both perturbed and unperturbed cloudless model spectra with \teff=500 K, \logg = 5.0 (cm s$^{-2}$) and the results are shown in Figure \ref{fig:hotspot}.

Perturbations at nearly all depths produce semi-amplitudes that are too large when compared to our observations.  Although perturbations at the 0.1 and 0.3 bar level of the atmosphere produced semi-amplitudes that are consistent with our [4.5] observations, they produce variability that is much too large at [3.6].  More troubling, however, is the fact that the hot spot models predict that the [3.6] amplitude is always larger than [4.5] amplitude, i.e. $A_{[3.6]}/A_{[4.5]} > >1$.  Given that our first epoch observations have a ratio less than unity and our second epoch observations have a ratio approximately equal to unity, hot spots can probably be ruled out as the underlying cause of the observed variability in \wise.

There are, however, two caveats to this assertion.  First, the amount of energy injected at a given layer of the atmosphere was arbitrarily set to 1.5 times the baseline emergent flux for illustrative purposes. Values less than 1.5 produce lower amplitudes that are more inline with our observations, but $A_{[3.6]}/A_{[4.5]}$ always remains greater than unity.  Second, we have only performed the calculation with a single, cloudless model.  Before hot spots can be definitively ruled out as a possible contributor to the variability in cool brown dwarfs, models at different temperatures and surface gravities that include the formation of clouds must be inspected.

\section{Discussion}

Since the detection of variability in brown dwarfs has so far been limited to the L and T dwarfs, it is important to place our observations of \wise\ in context with these results.  \citet{crossfield1} complied a catalog of late-type M, L, and T dwarfs with measured rotation periods, $v \sin i$ values, or variability amplitudes.  This catalog does not, however, include the recent results of \citet{2015ApJ...799..154M} who searched forty-four L and T dwarfs (L3--T8) for variability in the mid-infrared with \textit{Spitzer}. We have therefore combined the \citeauthor{2015ApJ...799..154M} results with the \citeauthor{crossfield1} catalog and converted peak-to-peak amplitudes to semi-amplitudes when appropriate.

Only four L and two T dwarfs have measured rotation periods larger than 8.5 hours making \wise\ one of the slower rotating field brown dwarfs known.  Indeed at spectral types later than T3, \wise\ has the longest period measured to date.  It is, however, currently unclear whether this has physical significance (e.g., brown dwarfs spin down as they cool) or is simply a result of observational bias (e.g., most searches do not have the time baseline to detect such long periods).  The [3.6] and [4.5] epoch 2 semi-amplitudes of 3.5\% are also the largest mid-infrared amplitudes observed to date with 2MASS J22282889$-$4310262 (T6) coming in second at 2.8\% \citep{2015ApJ...799..154M}.  Of course this amplitude is still far below that seen in the red-optical and near-infrared where semi-amplitudes of 5--13\% have been reported for some L and T dwarfs \citep{artigau1,radigan1,2013A&amp;A...555L...5G}.

\citeauthor{2015ApJ...799..154M} also found that the maximum variability amplitude in both the [3.6] and [4.5] bands increases through the L and T spectral classes, although the trend beyond T3 is based on a single T dwarf (the aforementioned 2MASS J22282889$-$4310262).  They fit a relation to the upper envelope of the [3.6] peak-to-peak amplitudes that predicts a maximum semi-amplitude of 4.25\% for a Y0.5 dwarf.  The epoch 2 [3.6] semi-amplitude of 3.6\% falls below this prediction and therefore based on a solitary Y dwarf, the \citeauthor{2015ApJ...799..154M} relation appears to hold across the T/Y boundary.  The epoch 2 semi-amplitude ratio of $A_{[3.6]}/A_{[4.5]} \approx 1$ for \wise\ is also formally consistent with the mean value of 1.0 (with a standard deviation of 0.7) for L and T dwarfs measured by \citet{2015ApJ...799..154M}.  However without a detection of variability at [3.6] in epoch 1, it remains unclear just how low the amplitude ratio can be for this object.

Finally, although these first observations of Y dwarf variability have given us some constraints on the timescales over which variability occurs, we are fundamentally limited in what we can learn about their atmospheres for two reasons.  First, our two mid-infrared wavelengths probe a limited range of atmospheric pressures high in the atmosphere (see Figure \ref{fig:TP}).  In addition, these layers fall well above the expected major KCl and Na$_2$S cloud decks which means it is likely that our observations are not directly probing atmospheric layers with clouds.  A high water cloud could also be a candidate for the variability in the second epoch because it forms at these high atmospheric layers, but such an explanation also faces difficulties as our best fitting models are not cold enough to form an optically thick cloud \citep[see][]{morley1}.  Nevertheless variability arising from water clouds should be considered in further studies.  Simultaneous, multi-wavelength observations over a larger wavelength range (i.e. near- and mid-infrared observations) would allow us to study multiple layers of the atmosphere with and without clouds.  Second, the models we compare observations to are one-dimensional and static and thus any attempt to compare time-dependent phenomena to them is at some level ad hoc.  Two- or three-dimensional models that include time-dependent phenomena like vertical mixing, cloud formation, and thermal relaxation are therefore sorely needed.

\acknowledgements
 
We thank the anonymous referee for comments that improved the paper.  This publication makes use of data products from the Wide-field Infrared Survey Explorer, which is a joint project of the University of California, Los Angeles, and the Jet Propulsion Laboratory/California Institute of Technology, funded by the National Aeronautics and Space Administrations and is based [in part] on observations made with the \textit{Spitzer Space Telescope}, which is operated by the Jet Propulsion Laboratory, California Institute of Technology under a contract with NASA. Support for this work was provided by NASA through an award issued by JPL/Caltech. and the NASA/ESA Hubble Space Telescope, obtained at the Space Telescope Science Institute, which is operated by the Association of Universities for Research in Astronomy, Inc., under NASA contract NAS 5-26555. This research has made use of the NASA/IPAC Infrared Science Archive, which is operated by the Jet Propulsion Laboratory, California Institute of Technology, under contract with the National Aeronautics and Space Administration.

\bibliographystyle{apj}
\bibliography{ydwarf}

\begin{thebibliography}{40}
\expandafter\ifx\csname natexlab\endcsname\relax\def\natexlab#1{#1}\fi

\bibitem[{{Ackerman} \& {Marley}(2001)}]{2001ApJ...556..872A}
{Ackerman}, A.~S., \& {Marley}, M.~S. 2001, \apj, 556, 872

\bibitem[{{Artigau} {et~al.}(2009){Artigau}, {Bouchard}, {Doyon}, \&
  {Lafreni{\`e}re}}]{artigau1}
{Artigau}, {\'E}., {Bouchard}, S., {Doyon}, R., \& {Lafreni{\`e}re}, D. 2009,
  \apj, 701, 1534

\bibitem[{{Bailer-Jones} \& {Mundt}(2001)}]{bailer-jones1}
{Bailer-Jones}, C.~A.~L., \& {Mundt}, R. 2001, \aap, 367, 218

\bibitem[{{Beuzit} {et~al.}(2008){Beuzit}, {Feldt}, {Dohlen}, {Mouillet},
  {Puget}, {Wildi}, {Abe}, {Antichi}, {Baruffolo}, {Baudoz}, {Boccaletti},
  {Carbillet}, {Charton}, {Claudi}, {Downing}, {Fabron}, {Feautrier},
  {Fedrigo}, {Fusco}, {Gach}, {Gratton}, {Henning}, {Hubin}, {Joos}, {Kasper},
  {Langlois}, {Lenzen}, {Moutou}, {Pavlov}, {Petit}, {Pragt}, {Rabou}, {Rigal},
  {Roelfsema}, {Rousset}, {Saisse}, {Schmid}, {Stadler}, {Thalmann}, {Turatto},
  {Udry}, {Vakili}, \& {Waters}}]{2008SPIE.7014E..18B}
{Beuzit}, J.-L., {et~al.} 2008, in Society of Photo-Optical Instrumentation
  Engineers (SPIE) Conference Series, Vol. 7014, Society of Photo-Optical
  Instrumentation Engineers (SPIE) Conference Series, 18

\bibitem[{{Buenzli} {et~al.}(2012){Buenzli}, {Apai}, {Morley}, {Flateau},
  {Showman}, {Burrows}, {Marley}, {Lewis}, \& {Reid}}]{buenzli1}
{Buenzli}, E., {et~al.} 2012, \apjl, 760, L31

\bibitem[{{Burgasser} {et~al.}(2002){Burgasser}, {Marley}, {Ackerman},
  {Saumon}, {Lodders}, {Dahn}, {Harris}, \&
  {Kirkpatrick}}]{2002ApJ...571L.151B}
{Burgasser}, A.~J., {Marley}, M.~S., {Ackerman}, A.~S., {Saumon}, D.,
  {Lodders}, K., {Dahn}, C.~C., {Harris}, H.~C., \& {Kirkpatrick}, J.~D. 2002,
  \apjl, 571, L151

\bibitem[{{Chamberlain} \& {Hunten}(1987)}]{1987IGS....36.....C}
{Chamberlain}, J.~W., \& {Hunten}, D.~M. 1987, Orlando FL Academic Press Inc
  International Geophysics Series, 36

\bibitem[{{Cohen} {et~al.}(2003){Cohen}, {Wheaton}, \&
  {Megeath}}]{2003AJ....126.1090C}
{Cohen}, M., {Wheaton}, W.~A., \& {Megeath}, S.~T. 2003, \aj, 126, 1090

\bibitem[{{Crossfield}(2014)}]{crossfield1}
{Crossfield}, I.~J.~M. 2014, \aap, 566, A130

\bibitem[{{Cushing} {et~al.}(2011){Cushing}, {Kirkpatrick}, {Gelino},
  {Griffith}, {Skrutskie}, {Mainzer}, {Marsh}, {Beichman}, {Burgasser},
  {Prato}, {Simcoe}, {Marley}, {Saumon}, {Freedman}, {Eisenhardt}, \&
  {Wright}}]{cushing1}
{Cushing}, M.~C., {et~al.} 2011, \apj, 743, 50

\bibitem[{{Cushing} {et~al.}(2006){Cushing}, {Roellig}, {Marley}, {Saumon},
  {Leggett}, {Kirkpatrick}, {Wilson}, {Sloan}, {Mainzer}, {Van Cleve}, \&
  {Houck}}]{2006ApJ...648..614C}
---. 2006, \apj, 648, 614

\bibitem[{{Dorren}(1987)}]{1987ApJ...320..756D}
{Dorren}, J.~D. 1987, \apj, 320, 756

\bibitem[{{Dupuy} \& {Kraus}(2013)}]{2013Sci...341.1492D}
{Dupuy}, T.~J., \& {Kraus}, A.~L. 2013, Science, 341, 1492

\bibitem[{{Dupuy} {et~al.}(2013){Dupuy}, {Kraus}, \&
  {Liu}}]{2013AAS...22115830D}
{Dupuy}, T.~J., {Kraus}, A.~L., \& {Liu}, M.~C. 2013, in American Astronomical
  Society Meeting Abstracts, Vol. 221, American Astronomical Society Meeting
  Abstracts, \#158.30

\bibitem[{{Enoch} {et~al.}(2003){Enoch}, {Brown}, \& {Burgasser}}]{enoch1}
{Enoch}, M.~L., {Brown}, M.~E., \& {Burgasser}, A.~J. 2003, \aj, 126, 1006

\bibitem[{{Fazio} {et~al.}(2004){Fazio}, {Hora}, {Allen}, {Ashby}, {Barmby},
  {Deutsch}, {Huang}, {Kleiner}, {Marengo}, {Megeath}, {Melnick}, {Pahre},
  {Patten}, {Polizotti}, {Smith}, {Taylor}, {Wang}, {Willner}, {Hoffmann},
  {Pipher}, {Forrest}, {McMurty}, {McCreight}, {McKelvey}, {McMurray}, {Koch},
  {Moseley}, {Arendt}, {Mentzell}, {Marx}, {Losch}, {Mayman}, {Eichhorn},
  {Krebs}, {Jhabvala}, {Gezari}, {Fixsen}, {Flores}, {Shakoorzadeh}, {Jungo},
  {Hakun}, {Workman}, {Karpati}, {Kichak}, {Whitley}, {Mann}, {Tollestrup},
  {Eisenhardt}, {Stern}, {Gorjian}, {Bhattacharya}, {Carey}, {Nelson},
  {Glaccum}, {Lacy}, {Lowrance}, {Laine}, {Reach}, {Stauffer}, {Surace},
  {Wilson}, {Wright}, {Hoffman}, {Domingo}, \& {Cohen}}]{fazio1}
{Fazio}, G.~G., {et~al.} 2004, \apjs, 154, 10

\bibitem[{{Foreman-Mackey} {et~al.}(2013){Foreman-Mackey}, {Hogg}, {Lang}, \&
  {Goodman}}]{2013PASP..125..306F}
{Foreman-Mackey}, D., {Hogg}, D.~W., {Lang}, D., \& {Goodman}, J. 2013, \pasp,
  125, 306

\bibitem[{{Gelino} {et~al.}(2002){Gelino}, {Marley}, {Holtzman}, {Ackerman}, \&
  {Lodders}}]{gelino1}
{Gelino}, C.~R., {Marley}, M.~S., {Holtzman}, J.~A., {Ackerman}, A.~S., \&
  {Lodders}, K. 2002, \apj, 577, 433

\bibitem[{{Gillon} {et~al.}(2013){Gillon}, {Triaud}, {Jehin}, {Delrez},
  {Opitom}, {Magain}, {Lendl}, \& {Queloz}}]{2013A&amp;A...555L...5G}
{Gillon}, M., {Triaud}, A.~H.~M.~J., {Jehin}, E., {Delrez}, L., {Opitom}, C.,
  {Magain}, P., {Lendl}, M., \& {Queloz}, D. 2013, \aap, 555, L5

\bibitem[{{Heinze} {et~al.}(2013){Heinze}, {Metchev}, {Apai}, {Flateau},
  {Kurtev}, {Marley}, {Radigan}, {Burgasser}, {Artigau}, \&
  {Plavchan}}]{heinze1}
{Heinze}, A.~N., {et~al.} 2013, \apj, 767, 173

\bibitem[{{Hillenbrand} {et~al.}(2002){Hillenbrand}, {Foster}, {Persson}, \&
  {Matthews}}]{2002PASP..114..708H}
{Hillenbrand}, L.~A., {Foster}, J.~B., {Persson}, S.~E., \& {Matthews}, K.
  2002, \pasp, 114, 708

\bibitem[{{Hogg} {et~al.}(2010){Hogg}, {Bovy}, \& {Lang}}]{2010arXiv1008.4686H}
{Hogg}, D.~W., {Bovy}, J., \& {Lang}, D. 2010, ArXiv e-prints

\bibitem[{{Khandrika} {et~al.}(2013){Khandrika}, {Burgasser}, {Melis}, {Luk},
  {Bowsher}, \& {Swift}}]{khandrika1}
{Khandrika}, H., {Burgasser}, A.~J., {Melis}, C., {Luk}, C., {Bowsher}, E., \&
  {Swift}, B. 2013, \aj, 145, 71

\bibitem[{{Kirkpatrick} {et~al.}(2012){Kirkpatrick}, {Gelino}, {Cushing},
  {Mace}, {Griffith}, {Skrutskie}, {Marsh}, {Wright}, {Eisenhardt}, {McLean},
  {Mainzer}, {Burgasser}, {Tinney}, {Parker}, \&
  {Salter}}]{2012ApJ...753..156K}
{Kirkpatrick}, J.~D., {et~al.} 2012, \apj, 753, 156

\bibitem[{{Knutson} {et~al.}(2008){Knutson}, {Charbonneau}, {Allen}, {Burrows},
  \& {Megeath}}]{knutson1}
{Knutson}, H.~A., {Charbonneau}, D., {Allen}, L.~E., {Burrows}, A., \&
  {Megeath}, S.~T. 2008, \apj, 673, 526

\bibitem[{{Marley} {et~al.}(2010){Marley}, {Saumon}, \& {Goldblatt}}]{marley1}
{Marley}, M.~S., {Saumon}, D., \& {Goldblatt}, C. 2010, \apjl, 723, L117

\bibitem[{{McBride} {et~al.}(2011){McBride}, {Graham}, {Macintosh}, {Beckwith},
  {Marois}, {Poyneer}, \& {Wiktorowicz}}]{2011PASP..123..692M}
{McBride}, J., {Graham}, J.~R., {Macintosh}, B., {Beckwith}, S.~V.~W.,
  {Marois}, C., {Poyneer}, L.~A., \& {Wiktorowicz}, S.~J. 2011, \pasp, 123, 692

\bibitem[{{Metchev} {et~al.}(2015){Metchev}, {Heinze}, {Apai}, {Flateau},
  {Radigan}, {Burgasser}, {Marley}, {Artigau}, {Plavchan}, \&
  {Goldman}}]{2015ApJ...799..154M}
{Metchev}, S.~A., {et~al.} 2015, \apj, 799, 154

\bibitem[{{Mohanty} {et~al.}(2002){Mohanty}, {Basri}, {Shu}, {Allard}, \&
  {Chabrier}}]{2002ApJ...571..469M}
{Mohanty}, S., {Basri}, G., {Shu}, F., {Allard}, F., \& {Chabrier}, G. 2002,
  \apj, 571, 469

\bibitem[{{Morales-Calder{\'o}n} {et~al.}(2006){Morales-Calder{\'o}n},
  {Stauffer}, {Kirkpatrick}, {Carey}, {Gelino}, {Barrado y Navascu{\'e}s},
  {Rebull}, {Lowrance}, {Marley}, {Charbonneau}, {Patten}, {Megeath}, \&
  {Buzasi}}]{2006ApJ...653.1454M}
{Morales-Calder{\'o}n}, M., {et~al.} 2006, \apj, 653, 1454

\bibitem[{{Morley} {et~al.}(2012){Morley}, {Fortney}, {Marley}, {Visscher},
  {Saumon}, \& {Leggett}}]{2012ApJ...756..172M}
{Morley}, C.~V., {Fortney}, J.~J., {Marley}, M.~S., {Visscher}, C., {Saumon},
  D., \& {Leggett}, S.~K. 2012, \apj, 756, 172

\bibitem[{{Morley} {et~al.}(2014{\natexlab{a}}){Morley}, {Marley}, {Fortney},
  \& {Lupu}}]{morley2}
{Morley}, C.~V., {Marley}, M.~S., {Fortney}, J.~J., \& {Lupu}, R.
  2014{\natexlab{a}}, \apjl, 789, L14

\bibitem[{{Morley} {et~al.}(2014{\natexlab{b}}){Morley}, {Marley}, {Fortney},
  {Lupu}, {Saumon}, {Greene}, \& {Lodders}}]{2014ApJ...787...78M}
{Morley}, C.~V., {Marley}, M.~S., {Fortney}, J.~J., {Lupu}, R., {Saumon}, D.,
  {Greene}, T., \& {Lodders}, K. 2014{\natexlab{b}}, \apj, 787, 78

\bibitem[{{Morley} {et~al.}(2014{\natexlab{c}}){Morley}, {Marley}, {Fortney},
  {Lupu}, {Saumon}, {Greene}, \& {Lodders}}]{morley1}
---. 2014{\natexlab{c}}, \apj, 787, 78

\bibitem[{{Radigan} {et~al.}(2012){Radigan}, {Jayawardhana}, {Lafreni{\`e}re},
  {Artigau}, {Marley}, \& {Saumon}}]{radigan1}
{Radigan}, J., {Jayawardhana}, R., {Lafreni{\`e}re}, D., {Artigau}, {\'E}.,
  {Marley}, M., \& {Saumon}, D. 2012, \apj, 750, 105

\bibitem[{{Radigan} {et~al.}(2014){Radigan}, {Lafreni{\`e}re}, {Jayawardhana},
  \& {Artigau}}]{2014ApJ...793...75R}
{Radigan}, J., {Lafreni{\`e}re}, D., {Jayawardhana}, R., \& {Artigau}, E. 2014,
  \apj, 793, 75

\bibitem[{{Reach} {et~al.}(2005){Reach}, {Megeath}, {Cohen}, {Hora}, {Carey},
  {Surace}, {Willner}, {Barmby}, {Wilson}, {Glaccum}, {Lowrance}, {Marengo}, \&
  {Fazio}}]{reach1}
{Reach}, W.~T., {et~al.} 2005, \pasp, 117, 978

\bibitem[{{Robinson} \& {Marley}(2014)}]{robinson1}
{Robinson}, T.~D., \& {Marley}, M.~S. 2014, \apj, 785, 158

\bibitem[{{Schneider} {et~al.}(2015){Schneider}, {Cushing}, {Kirkpatrick},
  {Gelino}, {Mace}, {Wright}, {Eisenhardt}, {Skrutskie}, {Griffith}, \&
  {Marsh}}]{2015ApJ...804...92S}
{Schneider}, A.~C., {et~al.} 2015, \apj, 804, 92

\bibitem[{{Showman} \& {Kaspi}(2013)}]{showman1}
{Showman}, A.~P., \& {Kaspi}, Y. 2013, \apj, 776, 85

\end{thebibliography}

\clearpage

\clearpage

\clearpage

\clearpage

\clearpage

\clearpage

\clearpage

\clearpage

\end{document}